\documentclass{article}

\PassOptionsToPackage{numbers, compress}{natbib}
\usepackage[preprint]{neurips_2024}

\usepackage[utf8]{inputenc} % allow utf-8 input
\usepackage[T1]{fontenc}    % use 8-bit T1 fonts
\usepackage{hyperref}       % hyperlinks
\usepackage{url}            % simple URL typesetting
\usepackage{booktabs}       % professional-quality tables
\usepackage{amsfonts}       % blackboard math symbols
\usepackage{nicefrac}       % compact symbols for 1/2, etc.
\usepackage{microtype}      % microtypography
\usepackage{xcolor}         % colors
\usepackage{natbib}
\usepackage{graphicx}
\usepackage{amsmath}
\usepackage{multirow}
\usepackage{subcaption}
\usepackage{amssymb}

\usepackage{caption}
\usepackage{wrapfig}
\usepackage{lipsum} % 提供示例文本
\usepackage{placeins}
\usepackage{adjustbox}
\usepackage{authblk}
\title{Molecular topological deep learning for polymer property prediction}

\author{%
  Cong Shen$^{1}$, Yipeng Zhang$^{2}$, Fei Han$^{1}$ and Kelin Xia$^{2,*}$ \\
  $^{1}$ Department of Mathematics, National University of Singapore, Singapore, 119076, Singapore\\
  $^{2}$ School of Physical and Mathematical Sciences, Nanyang Technological University, Singapore, 637371,Singapore\\
  $^{*}$ Corresponding author\\
  \texttt{cshen@nus.edu.sg, yipeng001@e.ntu.edu.sg, mathanf@nus.edu.sg, xiakelin@ntu.edu.sg} \\
}

\begin{document}

\maketitle

\begin{abstract}
Accurate and efficient prediction of polymer properties is of key importance for polymer design. Traditional experimental tools and density function theory (DFT)-based simulations for polymer property evaluation, are both expensive and time-consuming. Recently, a gigantic amount of graph-based molecular models have emerged and demonstrated huge potential in molecular data analysis. Even with the great progresses, these models tend to ignore the high-order and mutliscale information within the data. In this paper, we develop molecular topological deep learning (Mol-TDL) for polymer property analysis. Our Mol-TDL incorporates both high-order interactions and multiscale properties into topological deep learning architecture. The key idea is to represent polymer molecules as a series of simplicial complices at different scales and build up simplical neural networks accordingly. The aggregated information from different scales provides a more accurate prediction of polymer molecular properties.
\end{abstract}

\section{Introduction}\label{sec1}

Polymers are indispensable for our daily life and continues to drive innovation across industries, from packaging to healthcare and beyond \cite{korley2021toward, coates2020chemical}. The development of functional polymers has promoted the usage of polymers to a new level, that is not only within industrial and agricultural production, but goes beyond to healthcare, i.e., treating cancer and genetic diseases \cite{banerjee2021polymer, worch2019stereochemical}. Advanced polymer design and discovery will be the solutions to long-time problems in food, energy, environment, resources and other fields. Polymer informatics \cite{batra2021emerging, audus2017polymer, adams2008engineering} is to study  polymer properties computational tools. Its combination with artificial intelligence (AI) models has demonstrated great potential in polymer property predictors \cite{kuenneth2021polymer, doan2020machine, aldeghi2022graph, chen2021predicting,barnett2020designing,kim2021polymer, kern2021design, gurnani2021polyg2g, wu2019machine}.

Polymer AI models can be classified into two general types, i.e., fingerprint-based machine learning \cite{le2012quantitative, huan2015accelerated, moriwaki2018mordred} and end-to-end deep learning \cite{kuenneth2023polybert,xu2023transpolymer}. The first type is to represent chemical structures and physical properties as fingerprints/descriptors, which are then combined with machine learning models, such as random forest, gradient boosting tree, neural network, etc \cite{le2012quantitative, huan2015accelerated, moriwaki2018mordred}. This type of model is effective and very suitable for small or median-sized dataset. The second type is deep learning models, which are free from handcraft features. In particular, the combination of sequence representations, for instance, simplified molecular-input line-entry system (SMILES), with deep learning models, such as BERT \cite{kenton2019bert}, RoBERTa \cite{liu2021robustly}, GPT \cite{brown2020language}, ELMo \cite{peters2018dissecting} and XLM \cite{conneau2019cross}, has begun to show enormous power \cite{kuenneth2023polybert,xu2023transpolymer}. The Transformer model and its variants have been used to predict the properties of polymers. Among them are TransPolymer \cite{xu2023transpolymer} and polyBERT \cite{kuenneth2023polybert}, which use Transformer and BERT models respectively to analyze polymer SMILES sequence information. Further, geometric deep learning models, in particular Graph Neural Networks (GNNs), have significantly improved the prediction of polymer properties by capturing molecular spatial structure information \cite{kenton2019bert, liu2021robustly, brown2020language, peters2018dissecting, conneau2019cross, kuenneth2023polybert, xu2023transpolymer, fang2022geometry, shen2023molecular, yu2021molecular, fang2022molecular, wang2022molecular, zeng2018graph, aldeghi2022graph, gurnani2023polymer}.

More recently, topological deep learning (TDL) has been proposed a promising tool for analyzing data with complicated topological structures \cite{hajijtopological, bodnar2023topological, hajij2020cell}. Different from traditional graph neural networks, TDL employs special data representations, including hypergraph, simplicial complexes, cellular complexes, and combinatorial complex, and novel message-passing modules (between simplices of same/different dimensions) \cite{hensel2021survey, papillon2023architectures}. It shows huge potential in the characterization of more complex molecular structures and interactions, and for a better prediction of molecular properties \cite{bodnar2022neural, giusti2023cell, feng2019hypergraph, kim2020hypergraph, bai2021hypergraph}.

Inspired by TDL, we propose molecular topological deep learning (Mol-TDL) for polymer property prediction. Mol-TDL models the polymer monomer molecule by a series of simplicial complexes at different scales (through a filtration process), which characterize both higher-order and multiscale interactions within the data. For each simplicial complex generated at a particular scale, a simplicial neural network is employed. These simplicial neural networks from different scales, for the same polymer monomer molecule, are then aggregated together for the prediction of polymer property. Further, we develop a multiscale topological contrastive learning model and use it for (self-supervised) pre-training of simplex based message passing. It has been shown that our Mol-TDL model can achieve the-state-of-the-art in predicting polymer properties on well-established benchmark datasets.
\section{Related Works}\label{rw}

\subsection{Graph-based models for polymer property prediction}
Graph-based models have been proven to be efficient and powerful in capturing molecular structure information and predicting molecular properties. Especially, the development of graph neural networks has enabled researchers to obtain more effective molecular representations and greatly improve the performance of molecular property prediction. Among them, GEM \cite{fang2022geometry}, Mol-GDL \cite{shen2023molecular}, HM-GNN \cite{yu2021molecular}, KCL \cite{fang2022molecular} and MolCLR \cite{wang2022molecular} are typical representatives of using GNNs to predict molecular properties. Recently, GNNs have been used for the polymer data analysis. For instance, Graph Convolutional Neural Networks (GCNN) has been used in the prediction of dielectric constant and energy bandgap of polymers \cite{zeng2018graph}. The wDMPNN model uses graph-based representation of polymer structure and weighted directed message passing architecture for polymer property prediction \cite{aldeghi2022graph}. polyGNN combines graph neural network, multitask learning and other advanced deep learning techniques for predicting polymers property \cite{gurnani2023polymer}.

\subsection{TDL-based models}
Topological Deep Learning (TDL) \cite{hajijtopological, bodnar2023topological} leverages novel topological tools to characterize data with complicated higher-order structures. Different from graph-based data representation, TDL uses topological representations from algebraic topology, including cell complexes \cite{hajij2020cell, bodnar2022neural,giusti2023cell}, simplicial complexes \cite{bodnar2023topological, schaub2022signal}, sheaves \cite{hansen2019toward, bodnar2021weisfeiler}, combinatorial complexes \cite{hajijtopological}, and hypergraphs \cite{feng2019hypergraph, kim2020hypergraph, bai2021hypergraph}, to model not only pair-wise interactions (as in graphs), but also many-body or higher-order interactions among three or more elements. In fact, these algebraic topology-based molecular representations have already achieved great success in molecular data analysis, including protein flexibility and dynamic analysis \cite{xia2014persistent, sverrisson2021fast}, drug design \cite{cang2017topologynet}, virus analysis \cite{chen2022persistent}, materials property analysis \cite{reiser2022graph, townsend2020representation}. Further, TDL uses a generalized message-passing mechanism thus enables the communication of information from simplices of different dimensions. In contrast to GNNs, where information is passing among nodes or edges, TDL allows information to propagate through any neighborhood relation \cite{hajij2020cell, roddenberry2021principled, bodnar2023topological, hajijtopological, yang2023convolutional}.
\section{Methods}\label{sec2}

\subsection{Multiscale topological representation for polymer molecule}
\subsubsection{Simplicial complex representation for polymer molecule}
A simplicial complex is the generalization of a graph into its higher-dimensional counterpart. The simplicial complex is composed of simplexes. Each simplex is a finite set of vertices and can be viewed geometrically as a point (0-simplex), an edge (1-simplex), a triangle (2-simplex), a tetrahedron (3-simplex), and their k-dimensional counterpart ($k$-simplex). More specifically, a $k$-simplex $\sigma^k=\left\{v_0,v_1,v_2,\cdots, v_k\right\}$ is the convex hull formed by $k$ + 1 affinely independent points $v_0,v_1,v_2,\cdots, v_k$ as follows
\begin{equation} \nonumber
\sigma^{k} = \left\{ \lambda_{0}v_{0} + \lambda_{1}v_{1} + \cdots + \lambda_{k}v_{k} \middle| {\sum\limits_{i = 0}^{k}{\lambda_{i} = 1;\forall i,0 \leq \lambda_{i} \leq 1}} \right\}.
\end{equation}

The $i$th dimensional face of $\sigma^{k}(i<k)$ is the convex hull formed by $i$ + 1 vertices from the set of $k$ + 1 points $v_0,v_1,v_2,\cdots, v_k$. The simplexes are the basic components for a simplicial complex.

A simplicial complex $K$ is a finite set of simplexes that satisfy two conditions. First, any face of a simplex from $K$ is also in $K$. Second, the intersection of any two simplexes in $K$ is either empty or a shared face. A $k$th chain group $C_k$ is an Abelian group of oriented $k$-simplexes $\sigma^{k}$, which are simplexes together with an orientation, i.e., ordering of their vertex set. The boundary operator $\partial_{k}\left( C_{k}\rightarrow C_{k - 1} \right)$ for an oriented $k$-simplex $\partial_{k}$ can be denoted as
\begin{equation}\nonumber
\partial_{k}\sigma^{k} = {\sum\limits_{i = 0}^{k}{( - 1)^{i}\left\lbrack v_{0},v_{1},v_{2},\cdots,{\hat{v}}_{i},\cdots,v_{k} \right\rbrack}}.
\end{equation}

Here, $\left\lbrack v_{0},v_{1},v_{2},\cdots,{\hat{v}}_{i},\cdots,v_{k} \right\rbrack$ is an oriented ($k$ - 1)-simplex, which is generated by the original set of vertices except $v_i$. The boundary operator maps a simplex to its faces, and it guarantees that $\partial_{k-1}\partial_{k} = 0$. To facilitate a better description, we use notation $\sigma_{j}^{k - 1} \subset \sigma_{i}^{k}$ to indicate that $\sigma_{j}^{k - 1}$ is a face of $\sigma_{i}^{k}$. For two oriented $k$-simplexes, $\sigma_{i}^{k}$ and $\sigma_{j}^{k}$, of a simplicial complex $K$, they are upper adjacent, denoted as $\sigma_{i}^{k} \smallfrown \sigma_{j}^{k}$, if they are faces of a common ($k$ + 1)-simplex; they are lower adjacent, denoted as $\sigma_{i}^{k} \smallsmile \sigma_{j}^{k}$, if they share a common ($k$ - 1)-simplex as their face. The upper degree of a $k$-simplex $\sigma_{i}^{k}$, denoted as
$d_U(\sigma^{k})$, is the number of ($k$ + 1)-simplexes, of which $\sigma_{i}^{k}$ is a face. Similarly, we define the lower degree of $\sigma_{i}^{k}$, denoted as $d_L(\sigma^{k})$, to be the number of ($k$ - 1)-simplexes on the boundary of $\sigma_{i}^{k}$.

\subsubsection{Filtration-based multiscale representation}

A filtration process naturally generates a multiscale representation \cite{edelsbrunner2002topological}. The filtration parameter, which is key to the filtration process, is usually chosen as sphere radius (or diameter) for point cloud data, edge weight for graphs, and isovalue (or level set value) for density data. A systematical increase (or decrease) of the value for the filtration parameter will induce a sequence of hierarchical topological representations, which can be not only simplicial complexes but also graphs and hypergraphs. For instance, a filtration operation on a distance matrix, i.e., a matrix with distances between any two vertices as its entries, can be defined by using a cutoff value as the filtration parameter. More specifically, if the distance between two vertices is smaller than the cutoff value, an edge is formed between them. In this way, a systematical increase (or decrease) of the cutoff value will deliver a series of nested graphs, with the graph produced at a lower cutoff value as a part (or a subset) of the graph produced at a larger cutoff value. Similarly, nested simplicial complexes can be constructed by using various definitions of complexes, such as Vietoris-Rips complex, Čech complex, alpha complex, cubical complex, Morse complex, and clique complex. In this study, we employ the Vietoris-Rips complex to describe both the topological and geometric structure of a given point cloud. Formally, the Vietoris-Rips complex, denoted as $\mathrm{Rip}_r=\mathrm{Rip}_r(X)$, is defined for a point cloud $X$ and a cutoff value $r$, as follows:
\begin{equation} \nonumber
\mathrm{Rip}_r(X) := \left\{ \sigma \subseteq X \mid \sigma \text{ is finite and } \forall \mathbf{x}_i, \mathbf{x}_j \in \sigma, \; \Vert \mathbf{x}_i - \mathbf{x}_j \Vert \leq r \right\},
\end{equation}
where $\sigma$ represents a simplex in $X$, and $\Vert \mathbf{x}_i - \mathbf{x}_j \Vert$ denotes the Euclidean distance between any two points $\mathbf{x}_i$ and $\mathbf{x}_j$ in $\sigma$.

In Mol-TDL, pairwise distance between atoms is corresponding to the cutoff value $r$ in Vietoris-Rips complex $\mathrm{Rip}_r$. Given the significance of actual inter-atomic distances in our study, we set the parameter $r$ to span from $2.0$ to $4.0\, \text{\AA}$. To facilitate discretization within this range, we consider five distinct Vietoris-Rips complexes  $\mathrm{Rip}_r$, for $r=2.0$, $2.5$, $3.0$, $3.5$, and $4.0\, \text{\AA}$ in our model. Figure \ref{fig1} A shows an example of the filtration-based multiscale representation for a molecule. 

In the previous section, we discussed two $k$-simplex relations in a simplicial complex: the upper adjacency and lower adjacency relations. In this section, we consider these relations as interactions between $k$-simplexes within the mentioned Vietoris-Rips complexes $\mathrm{Rip}_r$. Specifically, we focus on the upper adjacency relation for all $0$-simplexes. This means an interaction occurs between two atoms if and only if their distance does not exceed $r$. This interaction is the same the traditional understanding of molecular graph interactions. When considering higher-order $k$-simplexes (where $k>0$), we conisder their lower adjacency; that is, two $k$-simplexes interact if and only if their common face is a $(k-1)$-simplex. We use adjacency matrix $A_{r,k}=\{A_{r,k}(i,j)\}$ to describe this interaction between $k$-simplexes in $\mathrm{Rip}_r$:

\begin{equation}\nonumber
A_{r,0}(i,j)=\left\{ \begin{matrix}
{1,~~\sigma_{i}^{r,0} \smallfrown \sigma_{j}^{r,0},i \neq j~~~~~~~} \\
{0,~~~~~~~~others.~~~~~~~~~~~~~}
\end{matrix} \right. 
\end{equation}

And for $k\geq 1:$
\begin{equation}\nonumber
A_{r,k}(i,j)=\left\{ \begin{matrix}
{1,~~\sigma_{i}^{r,k} \smallsmile \sigma_{j}^{r,k},i \neq j~~~~~~~} \\
{0,~~~~~~~~others.~~~~~~~~~~~~~}
\end{matrix} \right. 
\end{equation}
where $\sigma_{i}^{r,k}$ is the $i$-th $k$-simplex in $\mathrm{Rip}_r$. And the notation $\smallfrown$ and $\smallsmile$ represent the upper and lower adjacency relation, respectively.

Also we consider the following diagonal matrices $D_{r,k}=\{D_{r,k}(i,j)\}$ for normalization:

\begin{equation}\nonumber
D_{r,0}(i,j)=\left\{ \begin{matrix}
{d_U(\sigma_{i}^{r,0}),~~i = j~~~~~~} \\
{0,~~~~~~~~others.~~~~~~}
\end{matrix} \right. 
\end{equation}

And for $k\geq 1:$
\begin{equation}\nonumber
D_{r,k}(i,j)=\left\{ \begin{matrix}
{d_L(\sigma_{i}^{r,k}),~~i = j~~~~~~} \\
{0,~~~~~~~~others.~~~~~~}
\end{matrix} \right. 
\end{equation}
where $d_U(\sigma_{i}^{r,k})$ and $d_L(\sigma_{i}^{r,k})$ denote the upper and lower degree of $\sigma_{i}^{r,k}$, respectively.

\subsection{Multiscale topological deep learning}

\subsubsection{Simplex-based message passing}\label{sec:simplex}
In Mol-TDL, Vietoris-Rips complexes $\mathrm{Rip}_r$ is utilized to describe polymer. To be specific, we consider a series of adjacent matrices $A_{r,k}$ to describe the interaction between $k$-simplexes in $\mathrm{Rip}_r$. We use message passing to learn the feature representation of each simplex,
\begin{equation} \label{eq_GCN}
H_{r,k}^{(l + 1)} = {\rm Relu}\left({\hat{D}}_{r,k}^{- \frac{1}{2}}{\hat{A}}_{r,k}{\hat{D}}_{r,k}^{-\frac{1}{2}}H_{r,k}^{(l)}W_{r,k}^{(l)} \right).
\end{equation}

In the $l$-th iteration, the feature matrix $H_{r,k}^{(l + 1)}$ of $k$-simplexes is obtained by gathering neighbors feature of each $k$-simplex. Here $\hat{A}_{r,k}$ represents the sum of $A_{r,k}$ and identity matrix. $\hat{D}_{r,k}$ is a degree matrix, which is a diagonal matrix whose values on the diagonal are equal to the sum of the corresponding rows (or columns) in $\hat{A}_{r,k}$. $W_{r,k}^{(l)}$ is the weight matrix (to be learned). Computationally, we usually repeat the process 1 to 3 times, and the final simplex feature is denoted as $H_{r,k} $.

After message passing, the feature $H_{r,k}$ of $k$-simplex and its initial feature $H_{r,k}^{(0)}$ are concatenated together, and then all $k$-simplex features in $\mathrm{Rip}_r$ are gathered into one feature through a pooling process,
\begin{equation}\nonumber
f_{r,k} = {\rm Pooling} \left( [~H_{r,k}~||~H_{r,k}^{(0)}~]\right),
\end{equation}
where ${\rm Pooling}(\cdot)$ is a pooling function applied to all row vectors of the matrix and $[~\cdot~||~\cdot~]$ is concatenation operation.

In Mol-TDL, we consider five the filtration values. In this step, all features of $k$-simplexes in different filtration values are aggregated,
\begin{equation}\nonumber
f_{k} = {\rm READOUT}_{1}\left( f_{r,k}|~ r=2.0,~2.5,~\cdots,~ 4.0\right).
\end{equation}

Then all features in different orders are systematically aggregated,
\begin{equation}\nonumber
f = {\rm READOUT}_{2}\left( f_{k} ~ \middle|~ k=0,~1,~2 \right),
\end{equation}
where ${\rm READOUT}_1(\cdot)$ and ${\rm READOUT}_2(\cdot)$ both denote pooling function, and we choose concatenation in this study. Here $f$ is the final representation of a polymer.

Finally, a multiply layer perceptron (MLP) is utilized to property prediction,
\begin{equation}\nonumber
\hat{y} =  W^{1}{\rm Relu}\left( {W^{0}f + b^{0}} \right) + b^{1},
\end{equation}
where $W^{0}$ and $W^{1}$ are weight matrix, and $b^{0}$ and $b^{1}$ are the bias. The $\ell_1$ loss is implemented for regression tasks in Mol-TDL. Figure \ref{fig1}B  illustrates the flowchart of our Mol-TDL.

\subsubsection{Multiscale representation from simplex messages}
The initialization features of the $k$-simplex are crucial for predicting the properties of the polymer. Since the geometric meanings of $k$-simplex are obviously different, in Mol-TDL, different types of simplex have different feature initialization method.

For $0$-simplex, which represents atom in a molecule, we refer Mol-GDL \cite{shen2023molecular} and consider a total of 12 types of atoms, including C, H, O, N, P, Cl, F, Br, S, Si, I and all the rest atoms as one type. These atoms are chosen due to their high frequencies in the molecules in our datasets. For the $i$-th atom, atom type $a_j$ will contribute a component $g_i(r,\alpha_j)$ in the geometric node feature $g_i\left(r\right)=[g_i\left(r,\alpha_1\right),g_i\left(r,\alpha_2\right),\ldots,g_i(r,\alpha_{12})]$. Here $g_i(r,\alpha_j)$ means the number (or frequency) of all the neighboring atoms of type $\alpha_j$ for the $i$-th atom. For more details, please refer to Mol-GDL \cite{shen2023molecular}.

For $1$-simplex, which represents a bond in a molecule, we take the distance information as its initial features and use radial basis functions (RBF) \cite{buhmann2000radial} to represent this feature $g_{\rm RBF}\left( d, \varepsilon ,c \right) = e^{- \varepsilon{\|{d - c}\|}^{2}}$, where $d$ is distance between two nodes in a $1$-simplex. Here, $\varepsilon$ is a hyper-parameter and we set $\varepsilon = 1$. $c$ is center node and its value is chosen from $\{0,~0.1,~0.2,~...,~0.9\}$. Thus, a 10-dimensional initial feature vector, [$g_{\rm RBF}\left( d,1,c \right)|~c=0,~0.1,~0.2,~...,~0.9$] of $1$-simplex can be obtained after inputting a distance value $d$.

For $2$-simplex, which represents a triangle within a molecule, we take into account geometric information including area of the triangle, three related angles, and gravity center of the triangle. First, Heron's formula is used to calculate the area of the triangle, $g_{\rm area} = \sqrt{p\left( p - e_{1} \right)\left( p - e_{2} \right)\left( p - e_{3} \right)}$, where $e_{1}$, $e_{2}$ and $e_{3}$ represent the length of three edges, respectively. $p$ is half the circumference of a triangle, that is $p = \frac{1}{2}\left( e_{1} + e_{2} + e_{3} \right)$. Then, the cosines of the three angles in the triangle is utilized as the second part of the 2-simplex initial feature, that is $\cos_{1}$, $\cos_{2}$ and $\cos_{3}$. Next, the center of gravity of the triangle is calculated by averaging three atom coordinates, denoted as ${\rm cen}_{1}$, ${\rm cen}_{2}$ and ${\rm cen}_{3}$. Finally, all of this geometric information is concatenated to form a 7-dimensional initial feature of the 2-simplex, which is $\lbrack g_{area}, \cos_{1},\cos_{2}, \cos_{3}, {\rm cen}_{1}, {\rm cen}_{2}, {\rm cen}_{3}\rbrack$.

\subsection{Multiscale topological contrastive learning}
%\subsubsection{\color{red}Pretrain with contrastive learning models based on %simplical complex}

Motivated by recent developments in contrastive learning models for molecular property prediction \cite{wang2022molecular,fang2022molecular,li2022geomgcl}, we develop a multiscale topological contrastive learning for (self-supervised) pre-training of simplex-based message passing. In topological contrastive learning, pre-training is performed through maximizing the agreement between two augmented views of the same simplical complex via a contrastive loss in the latent space. The framework consists of the following four major components:

(1) \textbf{Topological data augmentation.} In graph models, graph data augmentation methods include node dropping, edge perturbation, attribute masking and subgraph sampling \cite{xu2021infogcl}. In our Mol-TDL model, we introduce simplex-based attribute masking, which is to mask part of (initial) simplex  features to augment the data.  For a given polymer, it can be represented as a series of $\mathrm{Rip}_r$, which undergo topological data augmentations to obtain two correlated views $H_{r,k}^{a}$ and $H_{r,k}^{b}$. Here $H_{r,k}^{a}$ represents original initial features of the $k$-simplex at cutoff distance $r$, and $H_{r,k}^{b}$ is generated by attribute masking. More specifically, we replace 30\% of the elements in the initial simplex features with random values. In contrastive learning, the pair of $H_{r,k}^{a}$ and $H_{r,k}^{b}$ forms a positive pair.

In our simplicial complex contrastive framework, the augmented views are utilized for a downstream task of label prediction. Inspired by the work\cite{tian2020makes}, we define the optimality of the views based on the mutual information. The main idea is that the optimal augmented views should retain all the information from the input simplicial complex relevant to the label, and all the shared information between the views should only be task-relevant.

\textbf{Proposition 1. (Optimal Augmented Views)} For a downstream task \( T \) of prediction of a semantic label \( y \), the optimal views, \( {(H_{r,k}^a)}^* \), \( {(H_{r,k}^b)}^*\), generated from the input simplicial complex Rip\( _r \) are the solutions to the following optimization problem:
\begin{equation}\label{opt1}
  (v_1^*, v_2^*) = \arg \min_{v_1, v_2} I(v_1; v_2)  
\end{equation}
subject to
\begin{equation}\label{cond1}
  I(v_1; y) = I(v_2; y) = I(\text{Rip}_r; y),
\end{equation}
Here $I(X; Y) := \mathbb{E}_{p(x,y)} \left[ \log \frac{p(x,y)}{p(x)p(y)} \right]$ denotes the mutual information of random variables $X,Y;$ $p(x,y)$ is the joint probability density function of $X,Y$ and $p(x),p(y)$ are the marginal probability density functions of $X$ and $Y$ respectively. Proposition 1 reveals that the shared information shared between the optimal views is minimized (Equation (\ref{opt1})). The proof of this proposition is in the Appendix.

(2) \textbf{Simplex-based message passing encoder.} A simplex-based message passing encoder (defined in Section \ref{sec:simplex}) extracts simplicial complex representation vectors $f^a$ and $f^b$ as in Equation \eqref{eq:similarity} for augmented simplicial complex pairs $H_{r,k}^{a}$ and $H_{r,k}^{b}$.

(3) \textbf{Contrastive loss function.}
A contrastive loss function $\mathcal{L}( \cdot )$ is defined to enforce maximizing the consistency between positive pairs $f^a$ and $f^b$ compared with negative pairs. Here the normalized temperature-scaled cross entropy loss \cite{wu2018unsupervised,sohn2016improved} is utilized to calculate contrastive loss.

During pre-training of simplex-based message passing, a minibatch of $M_{batch}$ polymer data are randomly sampled and processed through contrastive learning, resulting in $2M_{batch}$ augmented polymer data and corresponding contrastive loss to optimize, where we denote $j$-th polymer data pair as $f^a_j$ and $f^b_j$ in the minibatch. Negative pairs are not explicitly sampled but generated from the other $2M_{batch}-1$  augmented polymer data within the same minibatch as in \cite{chen2017sampling}. Denoting the cosine similarity function as,
\begin{equation}\label{eq:similarity}
sim \left( f^a_j,f^b_j \right) = \frac{(f^a_j)^{T}f^b_j} { \left\| f^a_j \right\| \left\| f^b_j \right\|}.
\end{equation}

The normalized temperature-scaled cross entropy loss \cite{wu2018unsupervised,sohn2016improved} for the $j$-th positive pair is defined as:
\begin{equation}\nonumber
\mathcal{L}(f^a_j,f^b_j) = -log\left( \frac{exp\left( sim\left( f^a_j,f^b_{j} \right)/\tau \right)}{\sum_{j' \neq j}^{M_{batch}}{exp\left( sim\left( f^a_j,f^b_{j'} \right)/\tau \right)}} \right),
\end{equation}
where $\tau$ denotes the temperature parameter. The final loss is computed across all positive pairs in the minibatch, i.e., $\sum_j \mathcal{L}(f^a_j,f^b_j)$.

\begin{figure}[t]
\centering
\includegraphics[width=0.8\textwidth]{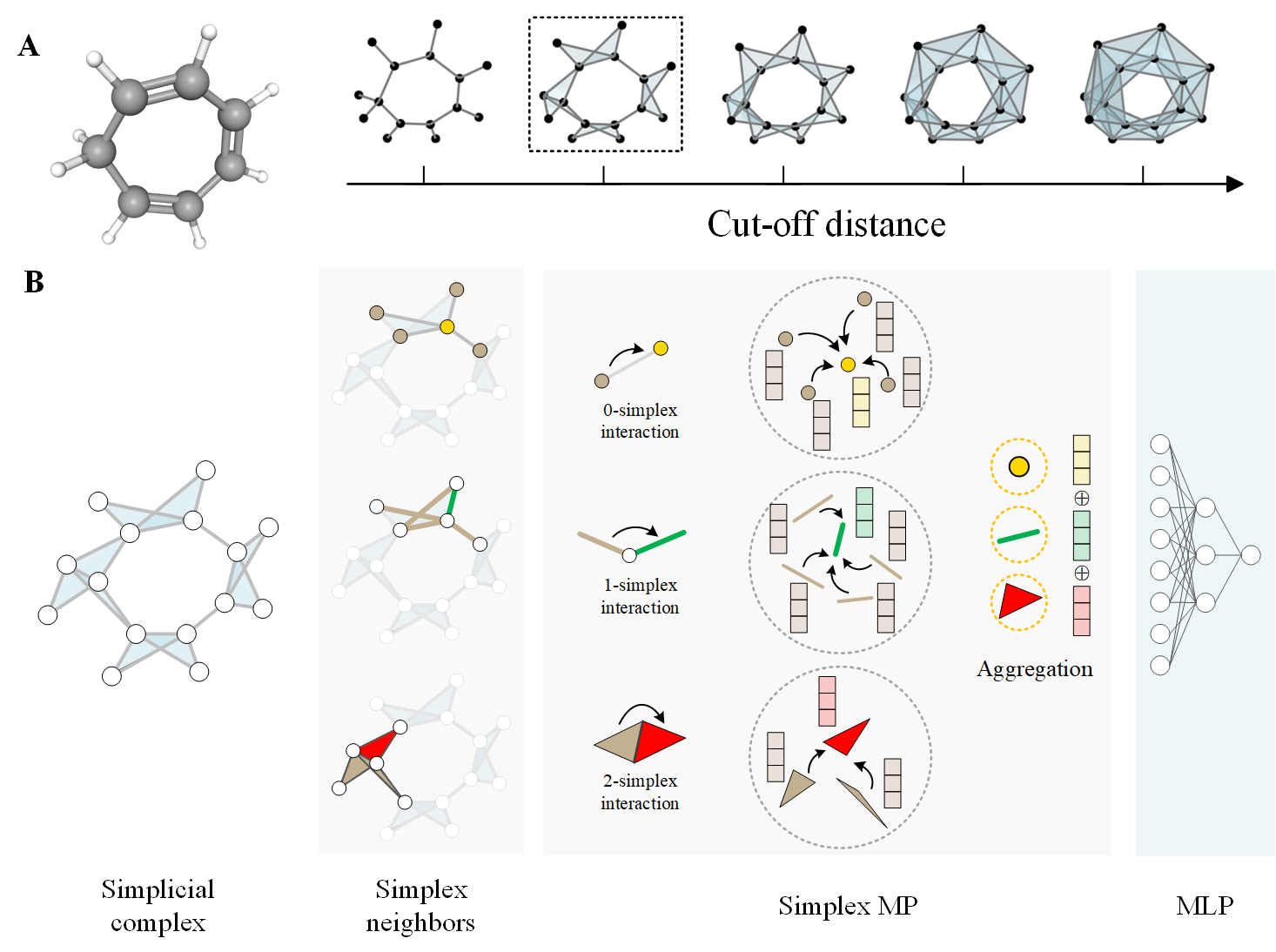}
\caption{Flowchart of Mol-TDL model. \textbf{A} Filtration of the Vietoris–Rips Complex for Cycloheptatriene. The Vietoris–Rips complex is applied at various cut-off distances: 2.0, 2.5, 3.0, 3.5, and 4.0 {\AA} in our model. \textbf{B} Topological deep learning model which utilizes a simplicial complex derived from the Vietoris–Rips complex with a cut-off distance. The neighborhood structure is constructed through interactions among 0, 1, and 2-simplexes. Message Passing (MP) in the model is implemented based on this neighborhood structure, followed by pooling operation and a Multilayer Perceptron (MLP) for regression analysis.}\label{fig1}
\end{figure}

\section{Results}
\subsection{Performance of Mol-TDL for polymer property prediction}

In this section, we present the comparison results between our Mol-TDL and SOTAs on different types of polymer data analysis. These datasets can be roughly classified into three categories: Electronic ($E_{gc}$\cite{kuenneth2021polymer}, $E_{gb}$\cite{kuenneth2021polymer}, $E_{ea}$\cite{kuenneth2021polymer}, $E_i$\cite{kuenneth2021polymer}, $E_{gap}^{crystal}$\cite{kamal2021novel}, $E_{gap}^{chain}$\cite{kamal2021novel}, $\Phi_e^{BC}$\cite{kamal2021novel}, ${OPV}_{Ave}$\cite{nagasawa2018computer}, ${OPV}_{Jsc}$\cite{nagasawa2018computer}, ${OPV}_{Voc}$\cite{nagasawa2018computer}, ${OPV}_{Eg}$\cite{nagasawa2018computer}, ${OPV}_{HOMO}$\cite{nagasawa2018computer}, ${OPV}_{LUMO}$)\cite{nagasawa2018computer}, Optical $\&$ dielectric($\varepsilon_0$\cite{kuenneth2021polymer}, $n_c$\cite{kuenneth2021polymer}) and Thermodynamic $\&$ physical($X_c$\cite{kuenneth2021polymer}). For these datasets, we only use their  repeat unit, i.e. monomers, to construct simplicial complexes and predict its performance. The detailed information of datasets can be found in Appendix Table \ref{tabs1}. All datasets of Mol-TDL model are split by ratio 8:1:1 for training, validation and test sets randomly. 

For benchmarking, the performance of Mol-TDL is compared with five state-of-the-art polymer/molecular property prediction models: polyBERT \cite{kuenneth2023polybert}, TransPolymer \cite{adams2008engineering}, polyGNN \cite{gurnani2023polymer}, Mol-GDL \cite{shen2023molecular} and GEM \cite{fang2022geometry}. Here, Root Mean Square Error (RMSE) and $R^2$ are used as metrics for evaluation in Mol-TDL and all benchmark. It is worth noting that we directly use the original results if the baseline model has RMSE and $R^2$ in the original literature, otherwise, we obtain the experimental results by running the code provided by the original literature.

Table \ref{tab1} shows that Mol-TDL can achieve good prediction performance on most data sets. Appendix Figure \ref{figs1}  and \ref{figs2} show the comparison between the predicted property with the density functional theory (DFT). Although the prediction models polyBERT and TransPolymer based on sequence information have larger-scale pre-training (polyBERT uses 100 million PSMILES strings, and TransPolymer uses 5 million augmented data for pre-training), Mol-TDL still has a competitive advantage even with a small pre-training scale. This shows that Mol-TDL has a more powerful advantage in capturing the spatial structure and topological information of polymers from the simplicial complex level. Compared with the classic models GEM and Mol-GDL used for molecular property prediction, Mol-TDL can still achieve prediction performance that is not inferior to these two models. GEM takes into account the correlation of higher-order interaction and other geometric information, like angles, but this model only learns its spatial structure information on the covalent bond molecular graph, while ignoring non-covalent bond information. Although the Mol-GDL model takes into account non-covalent bond information, it lacks more attention to high-order interaction and the geometric quantities in non-covalent bond information. Therefore, the performance of Mol-GDL in polymer property prediction is also inferior to Mol-TDL.

\begin{table}[h]
\caption{Performance of Mol-TDL and baseline models. (Note that * represents the results come from the original article, and N/A represents that the SMILES does not meet the model input specifications and no results can be obtained)}\label{tab1}%
\begin{adjustbox}{width=\textwidth}
\begin{tabular}{llllllll}
\toprule
&polyBERT & TransPolymer  & polyGNN & Mol-GDL & GEM & Mol-TDL\\
\midrule
\multicolumn{7}{l}{Metrics: $R^2$} \\
\midrule
$E_{gc}$   &$0.89_{\pm 0.02}^*$	&$\textbf{0.92}_{\pm 0.00}^*$   &$0.916_{\pm 0.005}^*$    &$0.869_{\pm 0.004}$ 	&$0.713_{\pm 0.018}$  &$0.881_{\pm 0.006}$\\
$E_{gb}$         &$0.93_{\pm 0.01}^*$	&$0.93_{\pm 0.01}^*$	&$0.839_{\pm 0.067}^*$	&$0.921_{\pm 0.005}$	&$0.860_{\pm 0.022}$  &$\textbf{0.936}_{\pm 0.005}$\\
$E_{ea}$         &$0.93_{\pm 0.03}^*$	&$0.91_{\pm 0.03}^*$	&$0.781_{\pm 0.108}^*$	&$0.596_{\pm 0.058}$	&$0.909_{\pm 0.032}$  &$\textbf{0.944}_{\pm 0.005}$ \\
$E_i$	          &$0.82_{\pm 0.07}^*$	&$0.84_{\pm 0.06}^*$	&$0.786_{\pm 0.156}^*$	&$0.758_{\pm 0.021}$	&$0.660_{\pm 0.035}$	&$\textbf{0.867}_{\pm 0.013}$  \\
$X_c$	          &$0.349_{\pm 0.016}$	&$0.50_{\pm 0.06}^*$	&$0.397_{\pm 0.073}^*$	&$0.421_{\pm 0.052}$	&$0.406_{\pm 0.093}$	&$\textbf{0.579}_{\pm 0.045}$ \\
$\varepsilon_0$  &$0.724_{\pm 0.012}$	&$0.76_{\pm 0.11}^*$	&$0.354_{\pm 0.404}^*$	&$0.716_{\pm 0.032}$	&$0.790_{\pm 0.043}$	&$\textbf{0.801}_{\pm 0.019}$  \\
$n_c$	                    &$0.86_{\pm 0.06}^*$	    &$0.82_{\pm 0.07}^*$	    &$0.543_{\pm 0.307}^*$	&$0.851_{\pm 0.012}$	&$0.828_{\pm 0.073}$	&$\textbf{0.882}_{\pm 0.021}$ \\
$E_{gap}^{crystal}$	    &$0.829_{\pm 0.005}$	&$0.750_{\pm 0.018}$	&N/A	&$0.819_{\pm 0.044}$	&$0.787_{\pm 0.135}$	&$\textbf{0.899}_{\pm 0.020}$ \\
$E_{gap}^{chain}$	        &$0.782_{\pm 0.004}$	&$0.861_{\pm 0.009}$	&N/A	&$0.844_{\pm 0.007}$	&$0.606_{\pm 0.031}$	&$\textbf{0.863}_{\pm 0.006}$  \\
$\Phi_e^{BC}$	&$0.650_{\pm 0.013}$	&$0.755_{\pm 0.006}$	&N/A	&$0.703_{\pm 0.019}$	&$\textbf{0.759}_{\pm 0.039}$	&$0.749_{\pm 0.014}$  \\
${OPV}_{Ave}$	&$0.299_{\pm 0.003}$	&$0.32_{\pm 0.05}^*$	    &N/A	&$0.350_{\pm 0.011}$	&$0.256_{\pm 0.022}$	&$\textbf{0.357}_{\pm 0.022}$  \\
${OPV}_{Jsc}$	&$0.294_{\pm 0.009}$	&$\textbf{0.354}_{\pm 0.023}$	&N/A	&$0.215_{\pm 0.016}$	&$0.194_{\pm 0.045}$	&$0.262_{\pm 0.028}$  \\
${OPV}_{Voc}$	&$0.241_{\pm 0.021}$	&$0.274_{\pm 0.019}$	&N/A	&$0.329_{\pm 0.038}$	&$0.305_{\pm 0.015}$	&$\textbf{0.503}_{\pm 0.043}$  \\
${OPV}_{Eg}$	&$0.401_{\pm 0.026}$	&$0.553_{\pm 0.033}$	&N/A	&$\textbf{0.569}_{\pm 0.032}$	&$0.356_{\pm 0.050}$ 	&$0.468_{\pm 0.036}$  \\
${OPV}_{HOMO}$	&$0.184_{\pm 0.035}$	&$\textbf{0.466}_{\pm 0.052}$	&N/A	&$0.401_{\pm 0.097}$	&$0.375_{\pm 0.087}$	&$0.379_{\pm 0.056}$  \\
${OPV}_{LUMO}$	&$0.375_{\pm 0.020}$	&$0.487_{\pm 0.034}$	&N/A	&$0.471_{\pm 0.027}$	&$0.133_{\pm 0.009}$	&$\textbf{0.514}_{\pm 0.035}$  \\
\midrule
\multicolumn{7}{l}{Metrics: $RMSE$} \\
\midrule
$E_{gc}$   &$0.701_{\pm 0.008}$	&$\textbf{0.44}_{\pm 0.01}^*$   &$0.442_{\pm 0.020}^*$    &$0.583_{\pm 0.009}$ 	&$0.560_{\pm 0.014}$  &$0.510_{\pm 0.012}$\\
$E_{gb}$         &$0.567_{\pm 0.020}$	&$0.52_{\pm 0.05}^*$	&$0.540_{\pm 0.170}^*$	&$0.553_{\pm 0.019}$	&$0.534_{\pm 0.042}$  &$\textbf{0.485}_{\pm 0.017}$\\
$E_{ea}$         &$0.308_{\pm 0.004}$	&$0.32_{\pm 0.02}^*$	&$0.341_{\pm 0.055}^*$	&$0.552_{\pm 0.596}$	&$0.274_{\pm 0.050}$  &$\textbf{0.263}_{\pm 0.013}$ \\
$E_i$	          &$0.525_{\pm 0.011}$	&$0.39_{\pm 0.07}^*$	&$0.540_{\pm 0.170}^*$	&$0.563_{\pm 0.024}$	&$\textbf{0.313}_{\pm 0.016}$	&$0.417_{\pm 0.021}$  \\
$X_c$	          &$17.646_{\pm 0.220}$	&$16.57_{\pm 0.68}^*$	&$18.6_{\pm 1.90}^*$	    &$18.768_{\pm 0.858}$	&$17.820_{\pm 1.684}$	&$\textbf{15.862}_{\pm 0.825}$ \\
$\varepsilon_0$  &$0.496_{\pm 0.011}$	&$0.52_{\pm 0.07}^*$	&$0.362_{\pm 0.086}^*$	&$0.501_{\pm 0.029}$	&$0.495_{\pm 0.066}$	&$\textbf{0.350}_{\pm 0.016}$  \\
$n_c$	                    &$0.131_{\pm 0.004}$	&$0.10_{\pm 0.02}^*$	    &$0.093_{\pm 0.030}^*$	&$0.086_{\pm 0.004}$	&$0.092_{\pm 0.023}$	&$\textbf{0.068}_{\pm 0.006}$ \\
$E_{gap}^{crystal}$	    &$0.613_{\pm 0.009}$	&$0.741_{\pm 0.027}$	&N/A	&$0.615_{\pm 0.071}$	&$0.702_{\pm 0.174}$	&$\textbf{0.567}_{\pm 0.057}$ \\
$E_{gap}^{chain}$	        &$0.683_{\pm 0.007}$	&$0.546_{\pm 0.018}$	&N/A	&$0.570_{\pm 0.013}$	&$0.777_{\pm 0.031}$	&$\textbf{0.539}_{\pm 0.012}$  \\
$\Phi_e^{BC}$	&$0.669_{\pm 0.013}$	&$0.560_{\pm 0.006}$	&N/A	&$0.581_{\pm 0.069}$	&$0.476_{\pm 0.038}$	&$\textbf{0.539}_{\pm 0.015}$  \\
${OPV}_{Ave}$	&$2.015_{\pm 0.005}$	&$1.92_{\pm 0.06}^*$	    &N/A	&$1.884_{\pm 0.016}$	&$2.267_{\pm 0.033}$	&$\textbf{1.843}_{\pm 0.031}$  \\
${OPV}_{Jsc}$	&$3.378_{\pm 0.022}$	&$\textbf{3.231}_{\pm 0.057}$	&N/A	&$3.740_{\pm 0.038}$	&$0.365_{\pm 0.025}$	&$3.644_{\pm 0.069}$  \\
${OPV}_{Voc}$	&$0.118_{\pm 0.002}$	&$0.115_{\pm 0.002}$	&N/A	&$0.129_{\pm 0.004}$	&$0.111_{\pm 0.001}$	&$\textbf{0.104}_{\pm 0.005}$  \\
${OPV}_{Eg}$	&$0.158_{\pm 0.003}$	&$0.137_{\pm 0.005}$	&N/A	&$\textbf{0.128}_{\pm 0.005}$	&$0.158_{\pm 0.006}$ 	&$0.146_{\pm 0.005}$  \\
${OPV}_{HOMO}$	&$0.209_{\pm 0.008}$	&$0.171_{\pm 0.008}$	&N/A	&$0.164_{\pm 0.008}$	&$0.168_{\pm 0.009}$	&$\textbf{0.161}_{\pm 0.007}$  \\
${OPV}_{LUMO}$	&$0.249_{\pm 0.004}$	&$0.209_{\pm 0.007}$	&N/A	&$0.211_{\pm 0.005}$	&$0.215_{\pm 0.001}$	&$\textbf{0.205}_{\pm 0.007}$  \\
\bottomrule

\end{tabular}
%\footnotetext{* represents the results come from the original article.}
%\footnotetext{N/A represents that the SMILES does not meet the model input specifications and no results can be obtained}
\end{adjustbox}
\end{table}
In order to further verify the Mol-TDL's representation ability for polymers, we compared the learned latent representations/spaces from Mol-TDL and polyBERT models. First, we visualize the pre-trained representations to verify whether they capture the scaffold information \cite{hu2016computational}, which is used to represent the core structures of bioactive compounds. Intuitively, molecules with the same scaffold share similar structure, and therefore are expected to be close in the high-level representation space. We randomly select four scaffolds and sample 100 molecules for each scaffold. Then we extract features of these molecules using Mol-TDL and polyBERT model. The $t$-distributed stochastic neighbor embedding ($t$-SNE) is used to reduce the dimensionality of the extracted features and visualize them into a 2D space. As shown in Appendix Figure \ref{figs3}A, it is not difficult to find that Mol-TDL and polyBERT are similar in their $t$-SNE distributions, where both can cluster different scaffolds well. However, the clustering effect of the Mol-TDL model is better than the polyBERT model on certain scaffold types, such as the scaffold represented by the blue points. Quantitatively, in terms of the Davies Bouldin index (DBI) \cite{davies1979cluster} (the smaller, the better), which is a metric to evaluate the clustering results, Mol-TDL clearly outperforms polyBERT model. Besides, We also randomly select three scaffolds and sample 100 molecules for each scaffold. The result is shown in Appendix Figure \ref{figs4}, and Mol-TDL also has better visualization performance. Another visualize test is that we use t-distributed stochastic neighbor embedding (t-SNE) to directly reduce the representation vectors learned by the Mol-TDL and polyBERT models to 2D space, as shown in Appendix Figure \ref{figs3}B and Appendix Figure \ref{figs5}. Overall, on these data sets, the Mol-TDL model's representation ability for polymers is significantly better than polyBERT. It is worth noting that on data sets such as $E_{gb}$, $E_{ea}$ and $E_{gap}^{chain}$, the advantages of the Mol-TDL model are particularly prominent because points of the same color are more obviously clustered together.

\subsection{Ablation study}
In Mol-TDL, instead of using only one molecular graph, a series of simplicial complexes are systematically generated by selecting different filtration cutoff value $r_{i}$ (see Method for details). Here, five filtration cutoff values, $r=2.0,2.5,3.0,3.5,4.0\, \text{\AA}$, are chosen to generate the simplicial complexes (Figure \ref{fig1} A) and three different order simplexes ($0$-simplex, $1$-simplex, $2$-simplex) are chosen for constructed higher-order interactions. The simplicial complex whose chemical bond length is between $0\, \text{\AA}$ and $2\, \text{\AA}$  is the $de facto$ standard of the covalent-bond-based model, and the other four are all constructed using covalent and non-covalent. For instance, a simplicial complex with a cut-off distance of  $3\, \text{\AA}$ means that the length of all the bond in this simplicial complex $k$ are no more than $3\, \text{\AA}$. Message passing in the model is implemented based on 0, 1, and 2-simplex interactions (See Figure \ref{fig1}  \textbf{B}). For example, when considering 2-simplex interaction, messages are passed between pairs of 2-simplexes that share this type of interaction.

Appendix Figure \ref{figs6}  shows the performance of Mol-TDL models under different cutoff distances and simplicial complices at different dimensions. More specifically, the "0-simplex" model means we only consider TDL based on 0-simplex message passing. The notation of "0,1-simplex" model means that both 0-simplex interactions and 1-simplex interactions are considered, while "0,1,2-simplex" model means that 0-simplex interactions, 1-simplex interactions and 2-simplex interactions are considered simultaneously. It can be seen that the consideration of higher order information can always improve the performance of the model, regardless of the cutoff distances. For both datasets, the performance of model will be enhanced when higher-order simplexes are added. The average values of the models from different cutoff distances are demonstrated in the right side of Appendix Figure \ref{figs6}. The same trend can be observed. Further, non-covalent interactions play an important role in the performance of models. In fact, the best performance is not at the cut-off distance of $2 \text{\AA}$, which usually indicates covalent interactions. Instead for the $\varepsilon_0$ dataset, the model has the best performance with a cut-off distance of $3\, \text{\AA}$, and the performance on the simplicial complexes with a cut-off distance of $3.5\, \text{\AA}$ and $4\, \text{\AA}$ are also better than that of the simplicial complex with a cut-off distance of $2\, \text{\AA}$. On the $E_{gb}$ data set, it can be seen more clearly that the prediction performance on simplicial complexx with a cut-off distance of $2\, \text{\AA}$ is worse than the performance on simplicial complexes with the other four higher cut-off distances.  The effects of non-covalent bonds and higher-order interactions on model performance in other data sets are shown in Appendix Table \ref{tabs2} to \ref{tabs9}.
 
In order to enable the Mol-TDL model to capture more structural information of polymers and further improve the performance of the model, multiscale topological contrastive learning and large-scale polymer structural data is used to pre-train Mol-TDL. We randomly select 1,000,000 polymers from TransPolymer \cite{adams2008engineering} and set the ratio of the training and validation set is 9:1 to pre-train Mol-TDL. Appendix Table \ref{tabs10} shows the difference in performance of the Mol-TDL model with and without pre-training. It is not difficult to find that the Mol-TDL model with pre-training is better than the Mol-TDL without pre-training as a whole. This result is more obviously reflected in $E_{ea}$, $E_i$ and $X_c$. However, it is worth noting that Mol-TDL pre-training does not improve performance on some data sets, such as $E_{gb}$ and $E_{gap}^{chain}$. We analyzed that the reason is that the amount of pre-training data is too small, and the $E_{gap}^{chain}$ used for fine-tuning are relatively large, so pre-training cannot provide more effective information for these fine-tuning data.
\section{Conclusion}
Efficient polymer representation learning is crucial for polymer property prediction. Existing works that apply graph neural network methods for polymer property prediction fail to characterize  higher-order and multiscale information. To this end, we design molecular topological deep learning (Mol-TDL), which incorporates both high-order interactions and multiscale properties into topological deep learning models. The key idea is to use a series of simplicial complexes at different scales to represent polymer monomer molecule and employ the simplex message-passing modules. In addition, multiscale topological contrastive learning is used to pre-train the model, thereby enriching the structural information of Mol-TDL from large-scale simplicial complexes. Our Mol-TDL model can achieve the state-of-the-art results in polymer benchmark datasets. 

\textbf{Limitations}  The Mol-TDL model can be further improved by the consideration of coupling of simplex message-passing modules at different dimensions. Currently, the simplex message-passing modules are done individually and lack communication between each other. A more sophisticated simplex message-passing architecture can be employed for more efficient characterization and learning of the high-order and multiscale information. 
\section*{Acknowledgments and Disclosure of Funding}

\begin{small}
\bibliographystyle{unsrt} % 或者 plain
\bibliography{refs} % 替换为您的参考文献文件名
\end{small}

%%%%%%%%%%%%%%%%%%%%%%%%%%%%%%%%%%%%%%%%%%%%%%%%%%%%%%%%%%%%
\newpage
\appendix

\section{Appendix / supplemental material}
\setcounter{figure}{0}
\setcounter{table}{0}
\setcounter{equation}{0}

\subsection{Proofs of corollaries}
\textbf{Lemma 1.} For any pair of view $(v_1, v_2)$ generated from Rips$_r$ for prediction of a semantic label $y$ that satisfies $I(v_1; y) = I(v_2; y) = I(\text{Rip}_r; y)$, the mutual information of the pair naturally has a lower bound: 
\[
  I(v_1; v_2) \geq I(\text{Rip}_r; y) 
\]

\textbf{Proof.} because the fact that $v_1,v_2$ are functions of Rip$_r$, we have $I(\text{Rip}_r; y)=I(v_1; y)\leq I(v_1, v_2; y)\leq I(\text{Rip}_r; y),$ which implies:
\[
  I(\text{Rip}_r; y)=I(v_1; y)= I(v_1, v_2; y)
\]
From the above Equation and $I(v_1, v_2; y)= I(v_1; y) + I(v_2; y\mid v_1),$ we know $I(v_2; y\mid v_1)=0.$ Thus, we have $I(v_1; v_2)=I(v_1; v_2)+I(v_2; y\mid v_1)=I(v_1,y;v_2)=I(v_2,y)+I(v_1,v_2|y)\geq I(v_2,y)=I(\text{Rip}_r; y).$

\textbf{Proposition 1. (Optimal Augmented Views)} For a downstream task \( T \) of prediction of a semantic label \( y \), the optimal views, \( {(H_{r,k}^a)}^* \), \( {(H_{r,k}^b)}^*\), generated from the input simplicial complex Rip\( _r \) are the solutions to the following optimization problem:
\begin{equation}\label{opt}
  (v_1^*, v_2^*) = \arg \min_{v_1, v_2} I(v_1; v_2)  
\end{equation}

subject to
\begin{equation}\label{cond1}
  I(v_1; y) = I(v_2; y) = I(\text{Rip}_r; y) 
\end{equation}

\textbf{Proof.} Because the optimal views \( {(H_{r,k}^a)}^* \), \( {(H_{r,k}^b)}^*\) retains all the information from the input simplicial complex Rip$_r$ relevant to the label $y$, the Equation (\ref{cond1}) holds for these views. Also, the shared information between the optimal views is only related to the label $y$, which means these views should be conditionally independent:
\begin{equation}\label{final}
  I({(H_{r,k}^a)}^*;{(H_{r,k}^b)}^*\mid y) = 0  
\end{equation}
From the proof of Lemma 1, we know the Equation (\ref{final}) implies $I({(H_{r,k}^a)}^*;{(H_{r,k}^b)}^*)$ reaches its lower bound $I(\text{Rip}_r; y).$ This proves that the optimal views ${(H_{r,k}^a)}^*$ and ${(H_{r,k}^b)}^*$ are solution of the optimal problem (\ref{opt}).

\newpage

\subsection{Supplementary figures and tables}

\begin{figure}[h]
\centering
\includegraphics[width=0.8\textwidth]{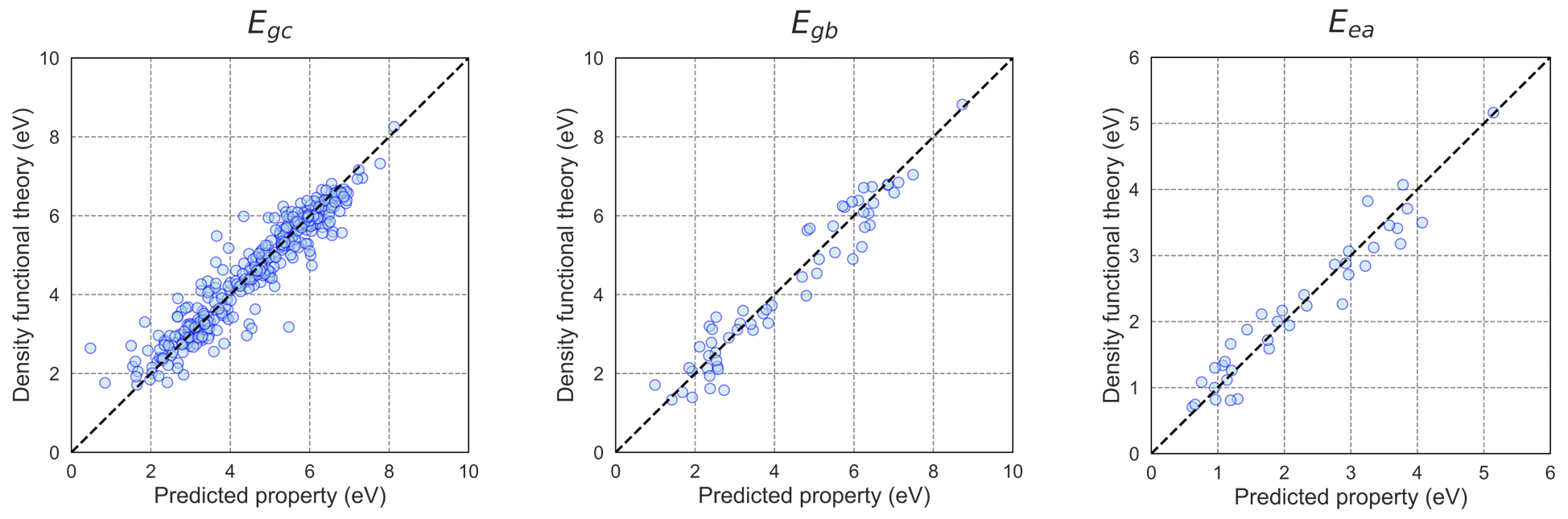}
\caption{Scatter plots of predicted values by Mol-TDL for three datasets: $E_{gc}$, $E_{gb}$, $E_{ea}$. The dashed lines on diagonals stand for perfect regression.}\label{figs1}
\end{figure}
\FloatBarrier

\begin{figure}[h]
\centering
\includegraphics[width=1.0\textwidth]{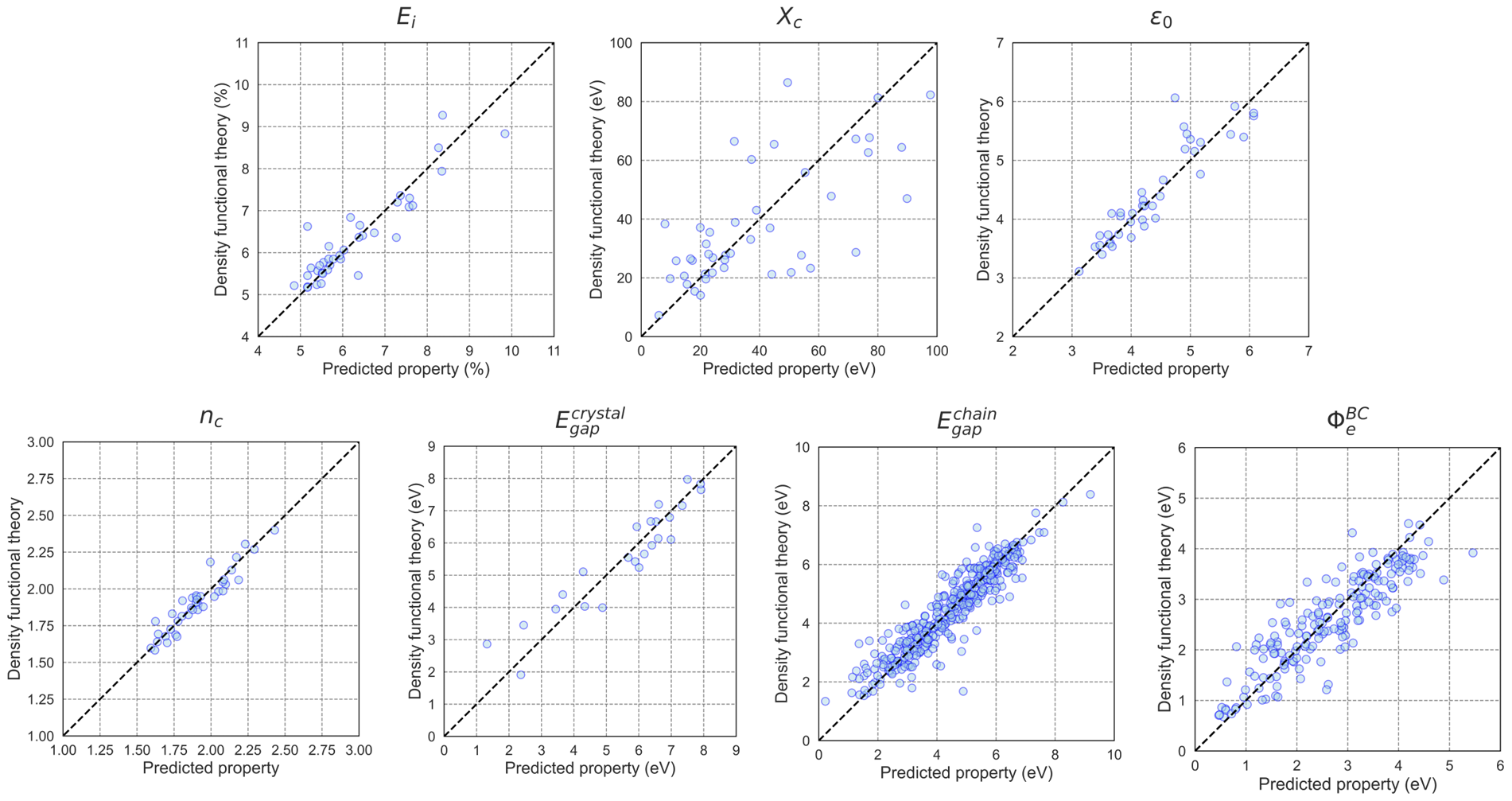}
\caption{ Scatter plots of predicted values by Mol-TDL for ten datasets: $E_i$, $X_c$, $\varepsilon_0$, $n_c$, $E_{gap}^{crystal}$, $E_{gap}^{chain}$, $\mathrm{\Phi}_e^{BC}$. The dashed lines on diagonals stand for perfect regression.}\label{figs2}
\end{figure}
\FloatBarrier

\begin{figure}[h]
\centering
\includegraphics[width=0.7\textwidth]{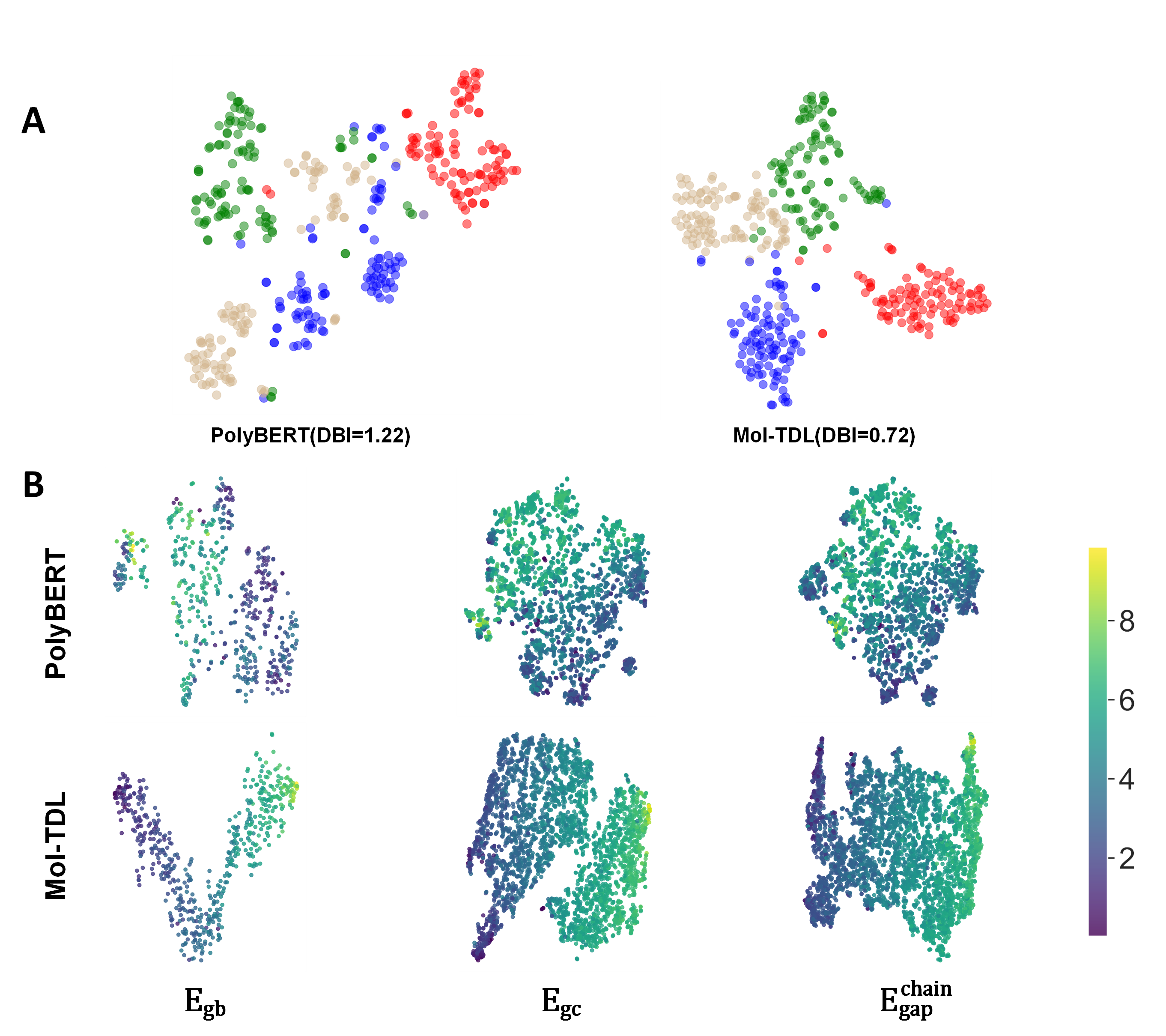}
\caption{Visualization of the representations learned by different models. \textbf{A.} Visualization of Mol-TDL and polyBERT latent representations of different scaffolds (indicated by different colors). \textbf{B.} Visualization of Mol-TDL and polyBERT latent representations for $E_{gb}$, $E_{ea}$ and $E_{gap}^{chain}$.}\label{figs3}
\end{figure}
\FloatBarrier

\begin{figure}[h]
\centering
\includegraphics[width=1.0\textwidth]{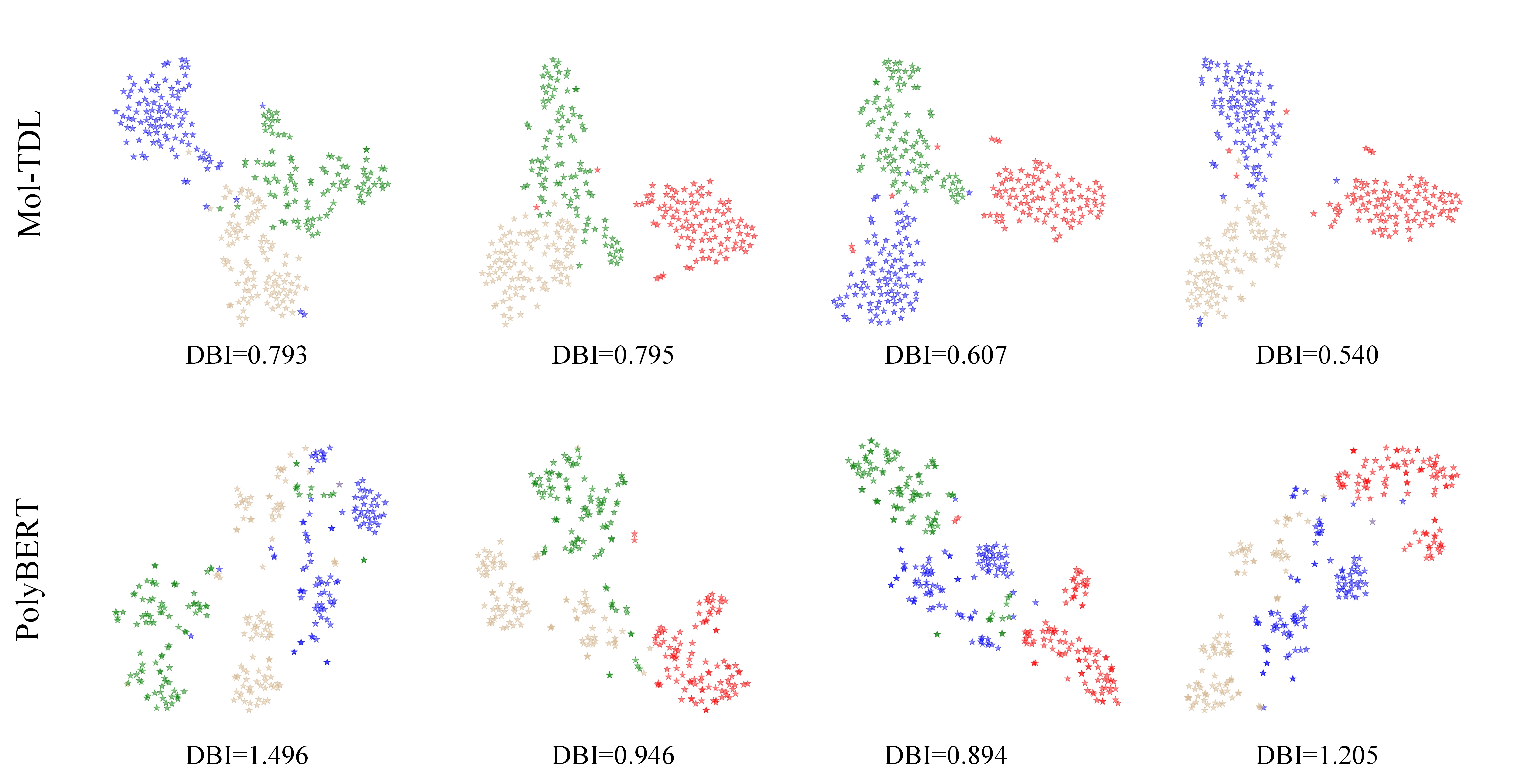}
\caption{The t-SNE visualization results with Mol-TDL and PolyBERT. The number in bracket indicates DBI and the smaller the number, the better the clustering effect. The different colors indicate different scaffolds.}\label{figs4}
\end{figure}
\FloatBarrier

\begin{figure}[h]
\centering
\includegraphics[width=1.0\textwidth]{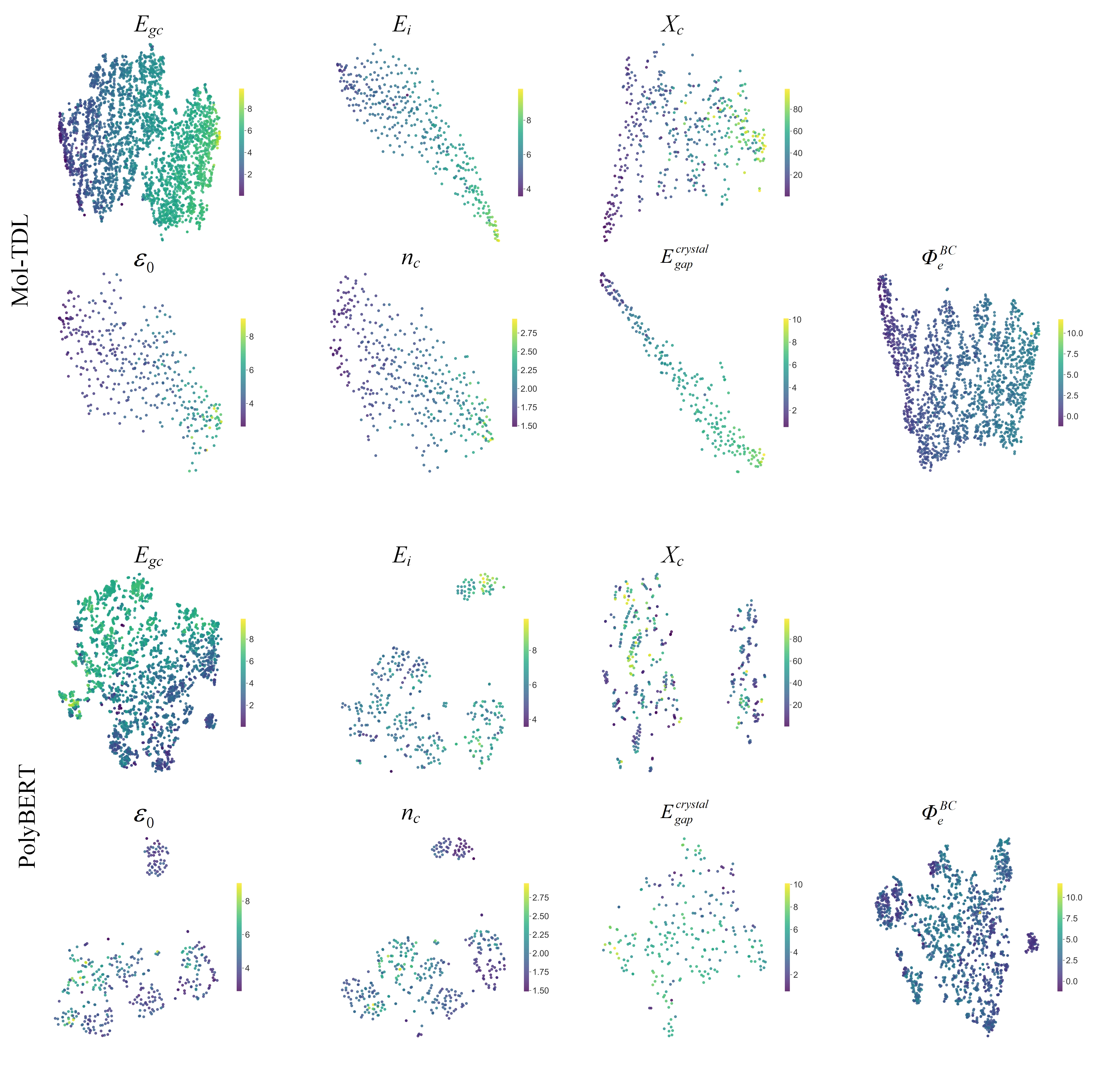}
\caption{Distribution visualization of predicted polymer property of Mol-TDL and PolyBERT for the seven test sets.}\label{figs5}
\end{figure}
\FloatBarrier

\begin{figure}[h]
\centering
\includegraphics[width=0.7\textwidth]{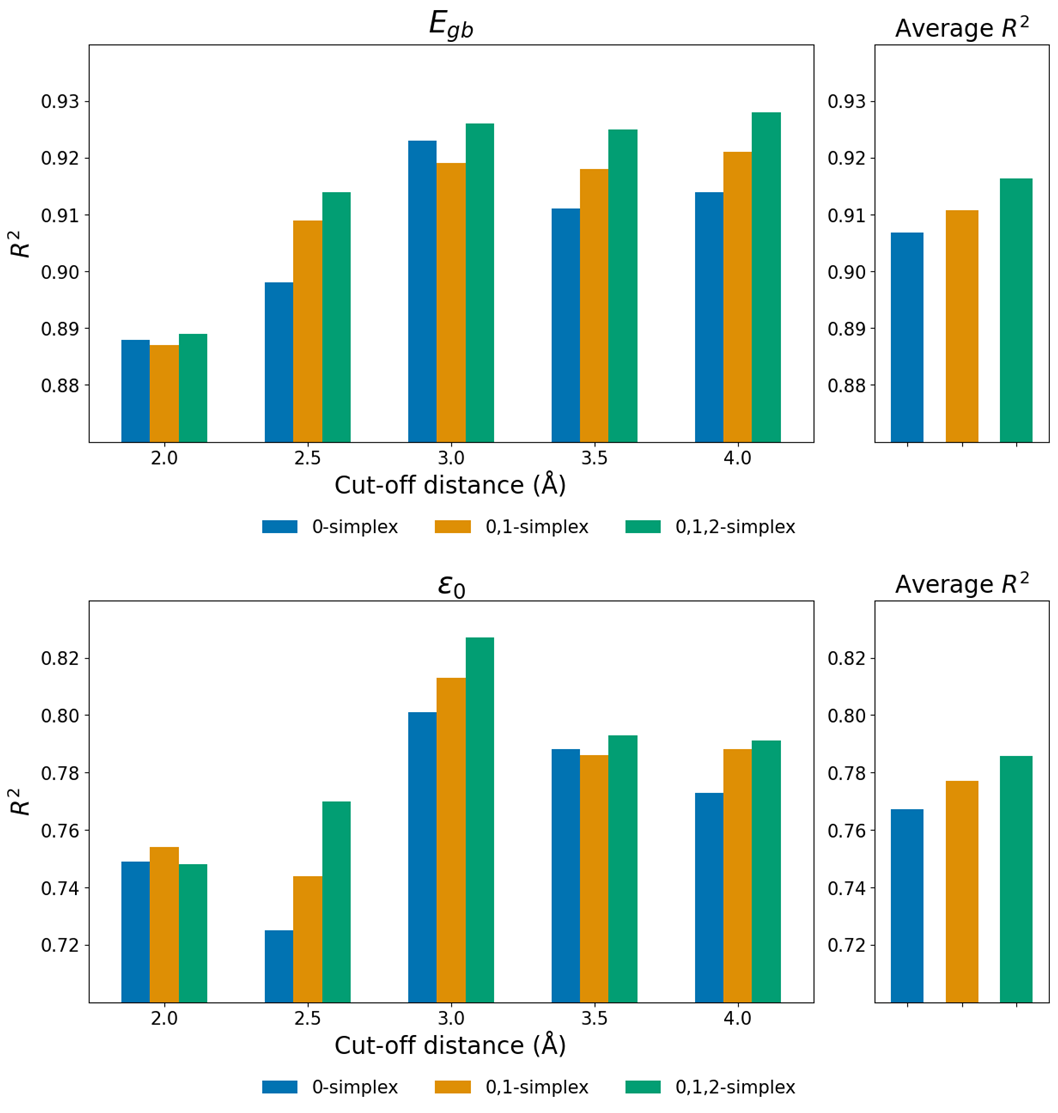}
\caption{The performance of polymer property prediction based on the Vietoris–Rips complex at various cut-off distances and different combination of higher-order simplex features.}\label{figs6}
\end{figure}
\FloatBarrier

\begin{table}[h]
\caption{Datasets for polymer property predictors}\label{tabs1}%
\begin{tabular}{llllllll}
\toprule
&Datasets &Property  &Units &Source &Data range &Polymers\\
\midrule
&$E_{gc}$   &bandgap (chain)	&eV   &DFT    &[0.02,9.86] 	&3390\\
&$E_{gb}$   &bandgap (bulk)	&eV   &DFT    &[0.4,10.1] 	&561\\
&$E_{ea}$   &electron affinity	&eV   &DFT    &[–0.39,5.17] 	&368 \\
&$E_i$	 &ionization energy	&eV   &DFT    &[3.56,9.84] 	&370  \\
&$X_c$	 &crystallization tendency	&\%   &DFT    &[0.1,98.8] 	&432\\
&$\varepsilon_0$  &dielectric constant	&1   &DFT    &[2.6,9.1] 	&382  \\
&$n_c$	&refractive index	&1   &DFT    &[1.48,2.95] 	&382\\
&$E_{gap}^{crystal}$	   &crystal bandgap	&eV   &DFT    &[0.5068,10.1137] 	&236 \\
&$E_{gap}^{chain}$	 &chain bandgap	&eV   &DFT    &[0.071,9.8351] 	&4209 \\
&$\Phi_e^{BC}$	&charge injection barrier	&eV   &DFT    &[-1.234,11.702] 	&1826 \\
&${OPV}_{Ave}$	&power conversion efficiency	&1   &DFT    &[0.01,10.5] 	&1203 \\
&${OPV}_{Jsc}$	&short-circuit current density	&$mA/cm^2$   &DFT    &[0.11,23.5] 	&1203  \\
&${OPV}_{Voc}$	&open-circuit voltage	&V   &DFT    &[0.2,1.09] 	&1203 \\
&${OPV}_{Eg}$	&bandgap	&eV   &DFT    &[1.0,2.4] 	&1203  \\
&${OPV}_{HOMO}$	&highest occupied molecular orbital	&eV   &DFT    &[4.24,6.18] 	&1203  \\
&${OPV}_{LUMO}$	&lowest unoccupied molecular orbital	&eV   &DFT    &[2.3,4.9] 	&1203  \\
\bottomrule

\end{tabular}\label{tabs1}
\end{table}
\FloatBarrier

\begin{table}[htbp]
\caption{The performance of polymer property prediction based on different interval molecular graph and different higher-order interactions (dataset: $E_{gc}$).}\label{tabs2}%
\begin{tabular}{lllllllll}
\toprule
\multirow{2}{*}{Simplex types}& \multicolumn{5}{l}{The cut-off distance}   &\multirow{2}{*}{Average}&  \\
\cmidrule(lr){2-6}
&2.0	\AA&2.5\AA& 3.0\AA&3.5\AA	&4.0\AA&\\
\midrule
\multicolumn{7}{l}{Metrics: $R^2$} \\
\midrule
0-simplex		&$0.842_{\pm0.032}$ 	&$0.851_{\pm0.023}$ 	&$0.836_{\pm0.008}$ 	&$0.836_{\pm0.013}$ 	&$0.855_{\pm0.011}$ 	&$0.844_{\pm0.009}$\\
					0,1-simplex	&$0.849_{\pm0.023}$ 	&$0.855_{\pm0.011}$ 	&$0.838_{\pm0.015}$ 	&$0.848_{\pm0.013}$ 	&$0.859_{\pm0.014}$ 	&$0.850_{\pm0.008}$\\
					0,1,2-simplex	&$0.854_{\pm0.018}$ 	&$0.872_{\pm0.015}$ 	&$0.854_{\pm0.024}$ 	&$0.853_{\pm0.010}$ 	&$0.865_{\pm0.004}$ 	&$0.860_{\pm0.009}$\\
\midrule
\multicolumn{7}{l}{Metrics: $RMSE$} \\
\midrule
0-simplex		&$0.585_{\pm0.061}$ 	&$0.568_{\pm0.044}$ 	&$0.598_{\pm0.014}$ 	&$0.597_{\pm0.024}$ 	&$0.562_{\pm0.021}$ 	&$0.582_{\pm0.016}$\\
					0,1-simplex	&$0.572_{\pm0.044}$ 	&$0.562_{\pm0.022}$ 	&$0.594_{\pm0.028}$ 	&$0.575_{\pm0.025}$ 	&$0.554_{\pm0.028}$ 	&$0.572_{\pm0.015}$\\
					0,1,2-simplex	&$0.564_{\pm0.034}$ 	&$0.527_{\pm0.030}$ 	&$0.562_{\pm0.048}$ 	&$0.566_{\pm0.020}$ 	&$0.543_{\pm0.009}$ 	&$0.552_{\pm0.017}$\\
\bottomrule

\end{tabular}\label{tabs2}
\end{table}
\FloatBarrier

\begin{table}[htbp]
\caption{The performance of polymer property prediction based on different interval molecular graph and different higher-order interactions (dataset: $E_{gb}$).}\label{tabs3}%
\begin{tabular}{lllllllll}
\toprule
\multirow{2}{*}{Simplex types}& \multicolumn{5}{l}{The cut-off distance}   &\multirow{2}{*}{Average}&  \\
\cmidrule(lr){2-6}
&2.0	\AA&2.5\AA& 3.0\AA&3.5\AA	&4.0\AA&\\
\midrule
\multicolumn{7}{l}{Metrics: $R^2$} \\
\midrule
0-simplex		&$0.888_{\pm0.005}$ 	&$0.898_{\pm0.021}$ 	&$0.923_{\pm0.008}$ 	&$0.911_{\pm0.006}$ 	&$0.914_{\pm0.006}$ 	&$0.907_{\pm0.014}$\\
					0,1-simplex	&$0.887_{\pm0.028}$ 	&$0.909_{\pm0.006}$ 	&$0.919_{\pm0.013}$ 	&$0.918_{\pm0.008}$ 	&$0.921_{\pm0.011}$ 	&$0.911_{\pm0.014}$\\
					0,1,2-simplex	&$0.889_{\pm0.019}$ 	&$0.914_{\pm0.009}$ 	&$0.926_{\pm0.009}$ 	&$0.925_{\pm0.007}$ 	&$0.928_{\pm0.012}$ 	&$0.916_{\pm0.016}$\\

\midrule
\multicolumn{7}{l}{Metrics: $RMSE$} \\
\midrule
0-simplex		&$0.643_{\pm0.015}$ 	&$0.610_{\pm0.062}$ 	&$0.532_{\pm0.029}$ 	&$0.573_{\pm0.020}$ 	&$0.563_{\pm0.020}$ 	&$0.584_{\pm0.043}$\\
					0,1-simplex	&$0.641_{\pm0.079}$ 	&$0.579_{\pm0.020}$ 	&$0.545_{\pm0.045}$ 	&$0.549_{\pm0.026}$ 	&$0.538_{\pm0.036}$ 	&$0.571_{\pm0.043}$\\
					0,1,2-simplex	&$0.637_{\pm0.055}$ 	&$0.564_{\pm0.030}$ 	&$0.522_{\pm0.031}$ 	&$0.526_{\pm0.025}$ 	&$0.513_{\pm0.043}$ 	&$0.552_{\pm0.051}$\\

\bottomrule

\end{tabular}\label{tabs3}
\end{table}
\FloatBarrier

\begin{table}[htbp]
\caption{The performance of polymer property prediction based on different interval molecular graph and different higher-order interactions (dataset: $E_{ea}$).}\label{tabs4}%
\begin{tabular}{lllllllll}
\toprule
\multirow{2}{*}{Simplex types}& \multicolumn{5}{l}{The cut-off distance}   &\multirow{2}{*}{Average}&  \\
\cmidrule(lr){2-6}
&2.0	\AA&2.5\AA& 3.0\AA&3.5\AA	&4.0\AA&\\
\midrule
\multicolumn{7}{l}{Metrics: $R^2$} \\
\midrule
0-simplex		&$0.935_{\pm0.007}$ 	&$0.919_{\pm0.012}$ 	&$0.920_{\pm0.006}$ 	&$0.928_{\pm0.011}$ 	&$0.923_{\pm0.016}$ 	&$0.929_{\pm0.015}$\\
					0,1-simplex	&$0.932_{\pm0.010}$ 	&$0.923_{\pm0.013}$ 	&$0.917_{\pm0.012}$ 	&$0.918_{\pm0.012}$ 	&$0.920_{\pm0.009}$ 	&$0.926_{\pm0.015}$\\
					0,1,2-simplex	&$0.924_{\pm0.006}$ 	&$0.904_{\pm0.019}$ 	&$0.920_{\pm0.019}$ 	&$0.929_{\pm0.008}$ 	&$0.912_{\pm0.011}$ 	&$0.922_{\pm0.015}$\\

\midrule
\multicolumn{7}{l}{Metrics: $RMSE$} \\
\midrule
0-simplex		&$0.246_{\pm0.021}$ 	&$0.330_{\pm0.025}$ 	&$0.330_{\pm0.012}$ 	&$0.312_{\pm0.023}$ 	&$0.323_{\pm0.034}$ 	&$0.308_{\pm0.036}$\\
					0,1-simplex	&$0.253_{\pm0.025}$ 	&$0.324_{\pm0.026}$ 	&$0.334_{\pm0.024}$ 	&$0.332_{\pm0.025}$ 	&$0.330_{\pm0.018}$ 	&$0.315_{\pm0.035}$\\
					0,1,2-simplex	&$0.276_{\pm0.016}$ 	&$0.359_{\pm0.035}$ 	&$0.328_{\pm0.042}$ 	&$0.310_{\pm0.019}$ 	&$0.345_{\pm0.022}$ 	&$0.324_{\pm0.032}$\\

\bottomrule

\end{tabular}\label{tabs4}
\end{table}
\FloatBarrier

\begin{table}[htbp]
\caption{The performance of polymer property prediction based on different interval molecular graph and different higher-order interactions (dataset: $E_{i}$).}\label{tabs5}%
\begin{tabular}{lllllllll}
\toprule
\multirow{2}{*}{Simplex types}& \multicolumn{5}{l}{The cut-off distance}   &\multirow{2}{*}{Average}&  \\
\cmidrule(lr){2-6}
&2.0	\AA&2.5\AA& 3.0\AA&3.5\AA	&4.0\AA&\\
\midrule
\multicolumn{7}{l}{Metrics: $R^2$} \\
\midrule
0-simplex		&$0.820_{\pm0.027}$ 	&$0.835_{\pm0.035}$ 	&$0.798_{\pm0.026}$ 	&$0.812_{\pm0.030}$ 	&$0.810_{\pm0.012}$ 	&$0.815_{\pm0.014}$\\
					0,1-simplex	&$0.816_{\pm0.030}$ 	&$0.856_{\pm0.009}$ 	&$0.820_{\pm0.022}$ 	&$0.828_{\pm0.014}$ 	&$0.815_{\pm0.025}$ 	&$0.827_{\pm0.017}$\\
					0,1,2-simplex	&$0.823_{\pm0.040}$ 	&$0.865_{\pm0.017}$ 	&$0.829_{\pm0.020}$ 	&$0.842_{\pm0.012}$ 	&$0.844_{\pm0.018}$ 	&$0.841_{\pm0.016}$\\

\midrule
\multicolumn{7}{l}{Metrics: $RMSE$} \\
\midrule
0-simplex		&$0.485_{\pm0.036}$ 	&$0.463_{\pm0.048}$ 	&$0.514_{\pm0.034}$ 	&$0.495_{\pm0.043}$ 	&$0.499_{\pm0.016}$ 	&$0.491_{\pm0.019}$\\
					0,1-simplex	&$0.490_{\pm0.040}$ 	&$0.434_{\pm0.014}$ 	&$0.485_{\pm0.030}$ 	&$0.475_{\pm0.020}$ 	&$0.492_{\pm0.033}$ 	&$0.475_{\pm0.024}$\\
					0,1,2-simplex	&$0.479_{\pm0.057}$ 	&$0.420_{\pm0.027}$ 	&$0.473_{\pm0.028}$ 	&$0.456_{\pm0.018}$ 	&$0.452_{\pm0.026}$ 	&$0.456_{\pm0.023}$\\

\bottomrule

\end{tabular}\label{tabs5}
\end{table}
\FloatBarrier

\begin{table}[htbp]
\caption{The performance of polymer property prediction based on different interval molecular graph and different higher-order interactions (dataset: $X_{c}$).}\label{tabs6}%
\begin{tabular}{lllllllll}
\toprule
\multirow{2}{*}{Simplex types}& \multicolumn{5}{l}{The cut-off distance}   &\multirow{2}{*}{Average}&  \\
\cmidrule(lr){2-6}
&2.0	\AA&2.5\AA& 3.0\AA&3.5\AA	&4.0\AA&\\
\midrule
\multicolumn{7}{l}{Metrics: $R^2$} \\
\midrule
0-simplex		&$0.475_{\pm0.095}$ 	&$0.441_{\pm0.112}$ 	&$0.387_{\pm0.098}$ 	&$0.459_{\pm0.081}$ 	&$0.357_{\pm0.083}$ 	&$0.424_{\pm0.050}$\\
					0,1-simplex	&$0.452_{\pm0.090}$ 	&$0.483_{\pm0.071}$ 	&$0.423_{\pm0.056}$ 	&$0.470_{\pm0.033}$ 	&$0.366_{\pm0.138}$ 	&$0.439_{\pm0.046}$\\
					0,1,2-simplex	&$0.486_{\pm0.041}$ 	&$0.362_{\pm0.088}$ 	&$0.363_{\pm0.061}$ 	&$0.384_{\pm0.090}$ 	&$0.380_{\pm0.050}$ 	&$0.395_{\pm0.052}$\\
 \midrule
\multicolumn{7}{l}{Metrics: $RMSE$} \\
\midrule
0-simplex		&$17.675_{\pm1.670}$ 	&$18.227_{\pm1.904}$ 	&$19.120_{\pm1.526}$ 	&$17.963_{\pm1.319}$ 	&$19.594_{\pm1.265}$ 	&$18.516_{\pm0.810}$\\
					0,1-simplex	&$18.076_{\pm1.484}$ 	&$17.569_{\pm1.227}$ 	&$18.577_{\pm0.908}$ 	&$17.820_{\pm0.563}$ 	&$19.387_{\pm2.176}$ 	&$18.286_{\pm0.720}$\\
					0,1,2-simplex	&$17.545_{\pm0.698}$ 	&$19.511_{\pm1.317}$ 	&$19.525_{\pm0.928}$ 	&$19.173_{\pm1.379}$ 	&$19.262_{\pm0.773}$ 	&$19.003_{\pm0.830}$\\

\bottomrule

\end{tabular}\label{tabs6}
\end{table}
\FloatBarrier

\begin{table}[htbp]
\caption{The performance of polymer property prediction based on different interval molecular graph and different higher-order interactions (dataset: $\varepsilon_0$).}\label{tabs7}%
\begin{tabular}{lllllllll}
\toprule
\multirow{2}{*}{Simplex types}& \multicolumn{5}{l}{The cut-off distance}   &\multirow{2}{*}{Average}&  \\
\cmidrule(lr){2-6}
&2.0	\AA&2.5\AA& 3.0\AA&3.5\AA	&4.0\AA&\\
\midrule
\multicolumn{7}{l}{Metrics: $R^2$} \\
\midrule
0-simplex		&$0.749_{\pm0.011}$ 	&$0.725_{\pm0.024}$ 	&$0.801_{\pm0.018}$ 	&$0.788_{\pm0.007}$ 	&$0.773_{\pm0.009}$ 	&$0.767_{\pm0.030}$\\
					0,1-simplex	&$0.754_{\pm0.006}$ 	&$0.744_{\pm0.025}$ 	&$0.813_{\pm0.011}$ 	&$0.786_{\pm0.020}$ 	&$0.788_{\pm0.015}$ 	&$0.777_{\pm0.028}$\\
					0,1,2-simplex	&$0.748_{\pm0.042}$ 	&$0.770_{\pm0.025}$ 	&$0.827_{\pm0.028}$ 	&$0.793_{\pm0.020}$ 	&$0.791_{\pm0.015}$ 	&$0.786_{\pm0.029}$\\

\midrule
\multicolumn{7}{l}{Metrics: $RMSE$} \\
\midrule
0-simplex		&$0.393_{\pm0.009}$ 	&$0.411_{\pm0.018}$ 	&$0.350_{\pm0.016}$ 	&$0.361_{\pm0.006}$ 	&$0.373_{\pm0.008}$ 	&$0.378_{\pm0.025}$\\
					0,1-simplex	&$0.389_{\pm0.005}$ 	&$0.397_{\pm0.019}$ 	&$0.339_{\pm0.010}$ 	&$0.362_{\pm0.018}$ 	&$0.361_{\pm0.013}$ 	&$0.369_{\pm0.023}$\\
					0,1,2-simplex	&$0.393_{\pm0.033}$ 	&$0.376_{\pm0.021}$ 	&$0.325_{\pm0.027}$ 	&$0.356_{\pm0.018}$ 	&$0.359_{\pm0.012}$ 	&$0.362_{\pm0.025}$\\

\bottomrule

\end{tabular}\label{tabs7}
\end{table}
\FloatBarrier

\begin{table}[htbp]
\caption{The performance of polymer property prediction based on different interval molecular graph and different higher-order interactions (dataset: $n_{c}$).}\label{tabs8}%
\begin{tabular}{lllllllll}
\toprule
\multirow{2}{*}{Simplex types}& \multicolumn{5}{l}{The cut-off distance}   &\multirow{2}{*}{Average}&  \\
\cmidrule(lr){2-6}
&2.0	\AA&2.5\AA& 3.0\AA&3.5\AA	&4.0\AA&\\
\midrule
\multicolumn{7}{l}{Metrics: $R^2$} \\
\midrule
0-simplex			&$0.869_{\pm0.013}$ 	&$0.876_{\pm0.015}$ 	&$0.892_{\pm0.006}$ 	&$0.886_{\pm0.006}$ 	&$0.893_{\pm0.010}$ 	&$0.883_{\pm0.011}$\\
					0,1-simplex		&$0.878_{\pm0.009}$ 	&$0.861_{\pm0.023}$ 	&$0.889_{\pm0.009}$ 	&$0.881_{\pm0.019}$ 	&$0.894_{\pm0.017}$ 	&$0.881_{\pm0.013}$\\
					0,1,2-simplex		&$0.899_{\pm0.013}$ 	&$0.873_{\pm0.019}$ 	&$0.895_{\pm0.011}$ 	&$0.882_{\pm0.012}$ 	&$0.894_{\pm0.012}$ 	&$0.889_{\pm0.011}$\\

\midrule
\multicolumn{7}{l}{Metrics: $RMSE$} \\
\midrule
0-simplex			&$0.072_{\pm0.004}$ 	&$0.070_{\pm0.004}$ 	&$0.065_{\pm0.002}$ 	&$0.067_{\pm0.002}$ 	&$0.065_{\pm0.003}$ 	&$0.068_{\pm0.003}$\\
					0,1-simplex		&$0.069_{\pm0.003}$ 	&$0.074_{\pm0.006}$ 	&$0.066_{\pm0.003}$ 	&$0.068_{\pm0.006}$ 	&$0.065_{\pm0.005}$ 	&$0.068_{\pm0.004}$\\
					0,1,2-simplex		&$0.063_{\pm0.004}$ 	&$0.071_{\pm0.005}$ 	&$0.064_{\pm0.003}$ 	&$0.068_{\pm0.003}$ 	&$0.064_{\pm0.004}$ 	&$0.066_{\pm0.003}$\\

\bottomrule

\end{tabular}\label{tabs8}
\end{table}
\FloatBarrier

\begin{table}[htbp]
\caption{The performance of polymer property prediction based on different interval molecular graph and different higher-order interactions (dataset: $E_{gap}^{crystal}$).}\label{tabs9}%
\begin{tabular}{lllllllll}
\toprule
\multirow{2}{*}{Simplex types}& \multicolumn{5}{l}{The cut-off distance}   &\multirow{2}{*}{Average}&  \\
\cmidrule(lr){2-6}
&2.0	\AA&2.5\AA& 3.0\AA&3.5\AA	&4.0\AA&\\
\midrule
\multicolumn{7}{l}{Metrics: $R^2$} \\
\midrule
0-simplex			&$0.851_{\pm0.025}$ 	&$0.781_{\pm0.068}$ 	&$0.798_{\pm0.060}$ 	&$0.748_{\pm0.067}$ 	&$0.796_{\pm0.020}$ 	&$0.795_{\pm0.037}$\\
					0,1-simplex		&$0.860_{\pm0.021}$ 	&$0.770_{\pm0.066}$ 	&$0.824_{\pm0.043}$ 	&$0.774_{\pm0.044}$ 	&$0.806_{\pm0.016}$ 	&$0.807_{\pm0.037}$\\
					0,1,2-simplex		&$0.808_{\pm0.021}$ 	&$0.813_{\pm0.052}$ 	&$0.865_{\pm0.014}$ 	&$0.811_{\pm0.012}$ 	&$0.812_{\pm0.018}$ 	&$0.822_{\pm0.024}$\\

\midrule
\multicolumn{7}{l}{Metrics: $RMSE$} \\
\midrule
0-simplex			&$0.688_{\pm0.059}$ 	&$0.828_{\pm0.137}$ 	&$0.797_{\pm0.115}$ 	&$0.892_{\pm0.114}$ 	&$0.807_{\pm0.041}$ 	&$0.802_{\pm0.074}$\\
					0,1-simplex		&$0.667_{\pm0.050}$ 	&$0.849_{\pm0.128}$ 	&$0.745_{\pm0.090}$ 	&$0.848_{\pm0.082}$ 	&$0.787_{\pm0.033}$ 	&$0.779_{\pm0.077}$\\
					0,1,2-simplex		&$0.783_{\pm0.044}$ 	&$0.767_{\pm0.116}$ 	&$0.656_{\pm0.035}$ 	&$0.777_{\pm0.024}$ 	&$0.775_{\pm0.038}$ 	&$0.752_{\pm0.054}$\\

\bottomrule

\end{tabular}\label{tabs9}
\end{table}
\FloatBarrier

\begin{table}[h]
\caption{Comparison of performance of Mol-TDL models with or without pre-training process.}\label{tabs10}%
\begin{tabular}{llllll}
\toprule
\multirow{2}{*}{Datasets}& \multicolumn{2}{l}{Without pre-training}   & \multicolumn{2}{l}{With pre-training}  \\
\cmidrule(lr){2-5}
&RMSE	&$R^2$	&RMSE	&$R^2$\\
\midrule
$E_{gc}$   &$0.516_{\pm 0.013}$ 	&$0.878_{\pm 0.006}$ 	&$\textbf{0.510}_{\pm 0.012}$	&$\textbf{0.881}_{\pm 0.006}$\\
$E_{gb}$   &$\textbf{0.485}_{\pm 0.017}$ 	&$\textbf{0.936}_{\pm 0.005}$ 	&$0.535_{\pm 0.058}$ 	&$0.922_{\pm 0.012}$ \\
$E_{ea}$   &$0.276_{\pm 0.008}$ 	&$0.934_{\pm 0.003}$ 	&$\textbf{0.263}_{\pm 0.013}$ 	&$\textbf{0.944}_{\pm 0.005}$ \\
$E_i$     &$0.464_{\pm 0.030 }$	&$0.835_{\pm 0.021}$ 	&$\textbf{0.417}_{\pm 0.021}$ 	&$\textbf{0.867}_{\pm 0.013}$ \\
$X_c$      &$16.448_{\pm 1.130}$	&$0.547_{\pm 0.063}$ 	&$\textbf{15.862}_{\pm 0.825}$ 	&$\textbf{0.579}_{\pm 0.045}$\\
$\varepsilon_0$ &$0.353_{\pm 0.017}$ 	&$0.797_{\pm 0.019}$ 	&$\textbf{0.350}_{\pm 0.016}$ 	&$\textbf{0.801}_{\pm 0.019}$ \\
$n_c$  &$0.069_{\pm 0.004}$ 	&$0.880_{\pm 0.013}$	&$\textbf{0.068}_{\pm 0.006}$	&$\textbf{0.882}_{\pm 0.021}$\\
$E_{gap}^{crystal}$ &$0.608_{\pm 0.034}$ 	&$0.884_{\pm 0.013}$ 	&$\textbf{0.567}_{\pm 0.057}$	&$\textbf{0.899}_{\pm 0.020}$\\
$E_{gap}^{chain}$   &$\textbf{0.539}_{\pm 0.012}$ 	&$\textbf{0.863}_{\pm 0.006}$ 	&$0.587_{\pm 0.010}$ 	&$0.837_{\pm 0.006}$ \\
$\Phi_e^{BC}$       &$0.541_{\pm 0.010}$ 	&$0.747_{\pm 0.010}$ 	&$\textbf{0.539}_{\pm 0.015}$ 	&$\textbf{0.749}_{\pm 0.014}$ \\
\bottomrule
\end{tabular}\label{tabs10}
\end{table}

%%%%%%%%%%%%%%%%%%%%%%%%%%%%%%%%%%%%%%%%%%%%%%%%%%%%%%%%%%%%

\end{document}